\documentstyle[12pt]{article}

\newcommand{\eq}{\begin{equation}}
\newcommand{\eqa}{\begin{eqnarray}}
\newcommand{\en}{\end{equation}}
\newcommand{\ena}{\end{eqnarray}}
\newcommand{\enn}{\nonumber \end{equation}}

\def\sk{\vskip .4cm}
\def\noi{\noindent}
\def\om{\omega}
\def\al{\alpha}
\def\be{\beta}

\def\Ga{\Gamma}

\def\la{q-q^{-1}}

\def\lam{{1 \over \la}}

\def\Cb{\bar{C}}

\def\epsi{\varepsilon}
\def\we{\wedge}
\def\ot{\otimes}

\def\de{\delta}

\def\part{\partial}

\def\R#1#2{ R^{#1}_{~~~#2} }
\def\PI#1#2{(P_I)^{#1}_{~~~#2} }
\def\PJ#1#2{ (P_J)^{#1}_{~~~#2} }
\def\Ppp{(P_+,P_+)}
\def\Ppm{(P_+,P_-)}
\def\Pmp{(P_-,P_+)}
\def\Pmm{(P_-,P_-)}
\def\Ppo{(P_+,P_0)}
\def\Pom{(P_0,P_-)}
\def\Pop{(P_0,P_+)}
\def\Pmo{(P_-,P_0)}
\def\Poo{(P_0,P_0)}
\def\Pso{(P_{\sigma},P_0)}
\def\Pos{(P_0,P_{\sigma})}

\def\Rh{{\hat R}}

\def\Rhat#1#2{ \Rh^{#1}_{~~~#2} }
\def\L#1#2{ \La^{#1}_{~~~#2} }
\def\Linv#1#2{ (\La^{-1})^{#1}_{~~~#2} }

\def\Rhatinv#1#2{ (\Rh^{-1})^{#1}_{~~~#2} }

\def\Z#1#2{ Z^{#1}_{~~~#2} }

\def\La{\Lambda}

\def\cchi#1#2{\chi^{#1}_{~#2}}
\def\ome#1#2{\om_{#1}^{~#2}}
\def\RRhat#1#2#3#4#5#6#7#8{\La^{~#2~#4}_{#1~#3}|^{#5~#7}_{~#6~#8}}

\def\LL#1#2#3#4#5#6#7#8{\La^{~#2~#4}_{#1~#3}|^{#5~#7}_{~#6~#8}}
\def\ZZ#1#2#3#4#5#6#7#8{Z^{~#2~#4}_{#1~#3}|^{#5~#7}_{~#6~#8}}
\def\WW#1#2#3#4#5#6#7#8{W^{~#2~#4}_{#1~#3}|^{#5~#7}_{~#6~#8}}
\def\PIJ#1#2#3#4#5#6#7#8{(P_I,P_J)^{~#2~#4}_{#1~#3}|^{#5~#7}_{~#6~#8}}

\def\Cb{{\bf C}}
\def\CC#1#2#3#4#5#6{\Cb_{~#2~#4}^{#1~#3}|_{#5}^{~#6}}

\def\T#1#2{ T^{#1}_{~~#2} }

\def\qm{q^{-1}}

\def\D{\Delta}

\def\qone{q \rightarrow 1}

\def\qonelim{\stackrel{q \rightarrow 1}{\longrightarrow}}

\def\ra{\rightarrow}

\def\BCD{B_n, C_n, D_n}

\begin{document}

\begin{titlepage}
\rightline{DFTT-18/93}
\vskip 2em
\begin{center}{\bf  A NOTE ON QUANTUM STRUCTURE CONSTANTS}
\\[6em]
 Leonardo Castellani ${}^{*}$
and Marco A. R-Monteiro${}^{\diamond *}$ \\[2em]
{\sl${}^{*}$Istituto Nazionale di
Fisica Nucleare, Sezione di Torino
\\and\\Dipartimento di Fisica Teorica\\
Via P. Giuria 1, 10125 Torino, Italy.}  \\
\sk

{\sl {}$^{\diamond}$Centro Brasileiro
 de Pesquisas Fisicas (CBPF)\\Rio de Janeiro, Brasil}\\
\vskip 2cm

\end{center}
\begin{abstract}
The Cartan-Maurer equations for any $q$-group
 of the $A_{n-1}, B_n, C_n, D_n$ series are given
 in a convenient  form, which allows their direct
 computation and  clarifies their connection with
 the $q=1$ case. These equations, defining the field
strengths,  are
essential in the construction of $q$-deformed gauge
 theories. An explicit expression
$\om ^i\we \om^j= -\Z {ij}{kl}\om ^k\we \om^l$ for
 the $q$-commutations of
left-invariant  one-forms is found, with
 $\Z{ij}{kl} \om^k \we \om^l  \qonelim \om^j\we\om^i$.
\end{abstract}

\vskip 5cm

\noi DFTT-18/93

\noi April 1993
\vskip .2cm
\noi \hrule
\vskip.2cm
\hbox{\vbox{\hbox{{\small{\it email addresses:}}}\hbox{}}
 \vbox{\hbox{{\small Decnet=31890::castellani;}}
\hbox{{\small Bitnet= castellani@to.infn.it }}}}

\end{titlepage}

\newpage
\setcounter{page}{1}

%\sect{Introduction }
Quantum groups \cite{Drinfeld}-\cite{Majid1} appear
as a  natural and consistent
algebraic structure behind continuously deformed physical
theories. Thus, in recent times,  there have been various
 proposals for deformed gauge theories and
gravity-like theories \cite{qgauge} based on $q$-groups.

Such deformations are interesting from different points of
 view, depending  also on which theory we are deforming. For
 example,  in quantized $q$-gravity
theories space-time becomes noncommutative, a fact
that does not contradict (Gedanken) experiments under
 the Planck length, and that could possibly provide a
regularization mechanism \cite{Connes,Majid2}.  On the
other hand, for the $q$-gauge theories constructed
 in \cite{Cas} spacetime can be taken to be  the ordinary
Minkowski spacetime,  the $q$-commutativity residing on
the fiber itself. As shown in \cite{Cas}, one can
construct a $q$-lagrangian invariant under $q$-gauge
 variations. This could suggest a way to break the classical
symmetry via a $q$-deformation, rather than by
introducing ad hoc scalar fields.  Note also that, unlike
 the $q=1$ case, the $q$-group $U_q(N)$ is simple, thus
 providing a ``quantum unification" of $SU(N) \otimes U(1)$.

In order to proceed from the algebraic $q$-structure to a
dynamical $q$-field theory, it is essential to investigate the
 differential calculus on $q$-groups. Indeed this provides
the $q$-analogues of the ``classical" definitions of
 curvatures, field strengths, exterior products of
 forms, Bianchi identities, covariant and Lie derivatives
and so on, see for ex.  \cite{Aschieri} for a review.
\sk
In this Letter we address and solve a specific problem: to find
 the Cartan-Maurer equations for any $q$-group of the $A,B,C,D$
 series in explicit form. These equations define the field strengths of
 the corresponding $q$-gauge theories \cite{Cas}.  The $A_{n-1}$ case
 was already treated in \cite{Aschieri}, where the structure constants
 were given explicitly, and shown to have the correct classical limit.

To our knowledge, this problem has been tackled
 previously only in ref. \cite{Watamura}. There, however, the
authors use (for the $B,C,D$ $q$-groups) a definition for
 the exterior product different
 from the one introduced in ref.s \cite{Wor}, adopted in
 \cite{Jurco,Zumino,Aschieri} and in
the present Letter.  As we will comment
later, their choice leads to a more complicated scenario.
\sk
Quantum groups are characterized by their $R$-matrix, which
 controls the noncommutativity of the quantum group
 basic elements $\T{a}{b}$ (fundamental representation):
\eq
\R{ab}{ef} \T{e}{c} \T{f}{d} = \T{b}{f} \T{a}{e} \R{ef}{cd} \label{RTT}
\en
and satisfies the quantum Yang-Baxter equation
\eq
\R{a_1b_1}{a_2b_2} \R{a_2c_1}{a_3c_2} \R{b_2c_2}{b_3c_3}=
\R{b_1c_1}{b_2c_2} \R{a_1c_2}{a_2c_3} \R{a_2b_2}{a_3b_3}, \label{QYB}
\en
a sufficient condition for the consistency of the
``RTT" relations (\ref{RTT}).  Its elements
depend
continuously on a (in general complex)
 parameter $q$, or even
on a set of parameters.  For $\qone$ we
 have $\R{ab}{cd} \qonelim \de^a_c \de^b_d$, i.e. the
 matrix entries $\T{a}{b}$ commute and
become the
 usual entries of the fundamental
 representation. The $q$-analogue
 of $\det T=1$, unitarity and orthogonality
 conditions can be
 imposed on the elements $\T{a}{b}$, consistently
with
the $RTT$ relations (\ref{RTT}), see \cite{FRT}.
\sk
The (uniparametric) $R$-matrices for the $q$-groups
of the $A_{n-1}, B_n, C_n, D_n$ series can be
found in ref. \cite{FRT}. We recall  the projector
 decomposition of the $\Rh$ matrix defined
by $\Rhat{ab}{cd} \equiv \R{ba}{cd}$, whose
$\qone$ limit is the permutation
operator $\de^a_d \de^b_c$ :
\sk
$A_n$ series:
\eq
\Rh=qP_+-\qm P_-  \label{RprojA}
\en
with
\eq
\begin{array}{ll}
&P_+={1 \over {q+\qm}} (\Rh+\qm I)\\
&P_-={1 \over {q+\qm}} (-\Rh+qI)\\
&I=P_++P_-
\end{array}
\label{projA}
\en
\sk
$B_n,C_n,D_n$ series:
\eq
\Rh=qP_+-\qm P_-+\epsi q^{\epsi-N}P_0  \label{RprojBCD}
\en
with
\eq
\begin{array}{ll}
&P_+={1 \over {q+\qm}} [\Rh+\qm I-(\qm+\epsi q^{\epsi-N})P_0]\\
&P_-={1 \over {q+\qm}} [-\Rh+qI-(q-\epsi q^{\epsi-N})P_0]\\
&P_0={{1-q^2} \over {(1-\epsi q^{N+1-\epsi})(1+\epsi q^{\epsi-N+1})}} K\\
&K^{ab}_{~~cd}=C^{ab} C_{cd}\\
&I=P_++P_-+P_0
\end{array}
\label{projBCD}
\en
where $\epsi=1$ for $B_n,D_n$, $\epsi=-1$ for $C_n$, and
$N$ is the dimension of the fundamental representation
$\T{a}{b}$ ($N=2n+1$ for $B_n$ and $N=2n$
 for $C_n,D_n$); $C_{ab}$  is the $q$-metric, and
$C^{ab}$ its inverse (cf. ref. \cite{FRT}).

{}From (\ref{RprojA}) and (\ref{RprojBCD}) we
 read off the eigenvalues of the $\Rh$ matrix, and
deduce the characteristic equations:
\eq
(\Rh-qI)(\Rh+\qm I)=0~~for~A_{n-1}~~(Hecke~condition) \label{Hecke}
\en
\eq
(\Rh-qI)(\Rh+\qm I)(\Rh-\epsi q ^{\epsi-N} I)=0,
{}~~for~B_n,C_n,D_n \label{cubic}
\en
\sk
The differential calculus on $q$-groups, initiated
 in ref.s \cite{Wor}, can
be entirely formulated in terms of the
 $R$ matrix. The
general constructive procedure can
be
found in  ref. \cite{Jurco}, or, in the
notations we adopt here, in
ref. \cite{Aschieri}.
\sk
As discussed in \cite{Wor} and
\cite{Jurco},
we can start by introducing
the (quantum) left-invariant one-forms
$\ome{a}{b}$, whose
exterior product
\eq
\ome{a_1}{a_2} \we \ome{d_1}{d_2}
\equiv \ome{a_1}{a_2} \otimes \ome{d_1}{d_2}
- \RRhat{a_1}{a_2}{d_1}{d_2}{c_1}{c_2}{b_1}{b_2}
\ome{c_1}{c_2} \otimes \ome{b_1}{b_2} \label{exteriorproduct}
\en
\noi is defined by the braiding matrix $\La$:
\eq
\LL{a_1}{a_2}{d_1}{d_2}{c_1}{c_2}{b_1}{b_2}
\equiv  d^{f_2} d^{-1}_{c_2}
\Rhat{b_1f_2}{c_2g_1} \Rhatinv{c_1g_1}{a_1e_1}
    \Rhatinv{a_2e_1}{d_1g_2}
\Rhat{d_2g_2}{b_2f_2} \label{Lambda}
\en
\noi For $q\ra 1$ the braiding
matrix $\Lambda$ becomes the usual
permutation
operator and one recovers the classical
exterior product.
Note that the ``quantum cotangent space" $\Ga$, i.e. the space
  spanned by the
quantum one-forms $\ome{a}{b}$, has dimension
$N^2$, in general bigger than its classical
counterpart ($dim\Ga=N^2$ only for the $U_q(N)$ groups).  This
is necessary in order to have a bicovariant bimodule structure
for $\Ga$ (cf. ref.  (\cite{Watamura}.). The same phenomenon occurs
 for the $q$-Lie generators defined below. For these, however, one
 finds restrictions (induced by the conditions imposed
 on the $\T{a}{b}$ elements) that  in general reduce
 the number of independent generators.
Working with $N^2$ generators is more convenient, since
 the nice quadratic relations (\ref{qLiealgebra}) of the
 $q$-Lie algebra become of higher order if one expresses
them in terms of a reduced set of independent
generators. For a discussion see \cite{Zumino}.
\sk
The relations (\ref{Hecke}) and (\ref{cubic}) satisfied
 by the $\Rh$ matrices
of the $A$ and $B,C,D$ series
 respectively reflect
 themselves in the relations
for the matrix $\La$:
\eq
(\La +q^2 I)(\La+q^{-2}I)(\La -I)=0  \label{Aspectral}
\en
for the $A$ $q$-groups, and
\eq
\begin{array}{l}
(\La +q^2 I)(\La+q^{-2}I)(\La+
\epsi q^{\epsi+1-N}I)(\La+\epsi q^{N-\epsi-1}I)
{}~~~~~~~~~~~\\
{}~~~~~~~~~~~~~~~~~~~~~~\times~(\La-\epsi q^{N+1-\epsi}I)
(\La-\epsi q^{-N-1+\epsi}I)(\La -I)
=0  \label{BCDspectral}
\end{array}
\en
for the $B,C,D$ $q$-groups, with the
same $\epsi$ as in (\ref{cubic}). We give later
an easy proof of these two relations.
\sk
Besides defining the exterior
product of forms, the matrix $\La$ contains
all the the information about
the quantum Lie algebra corresponding to the $q$-group.
\sk
The exterior differential of a quantum k-form $\theta$
 is defined by means of the
bi-invariant (i.e. left- and right-invariant) element  $\tau=\sum_a
\ome{a}{a} $ as follows:
\eq
d\theta \equiv \lam [\tau \we \theta - (-1)^k \theta \we \tau],
\label{exteriordifferential}
\en
The normalization $\lam$ is necessary in order to obtain the
correct classical limit (see for ex. \cite{Aschieri}). This linear
 map satisfies $d^2=0$, the Leibniz rule
and commutes with the left and right action
of the $q$-group \cite{Jurco}.
\sk
The exterior differentiation allows the definition of the
 ``quantum Lie algebra generators"
$\cchi{a_1}{a_2}$, via the formula \cite{Wor}
\eq
da=\lam[\tau a - a\tau] =(\cchi{a_1}{a_2} * a)
\ome{a_1}{a_2}. \label{qgenerators}
\en
\noi where
\eq
{}~~\chi * a \equiv (id \otimes \chi) \D (a),~~~~\forall a \in G_q,~
\chi \in G_q'
\en
and $\D$ is the usual coproduct on the quantum
 group $G_q$,  defined by
 $\D(\T{a}{b})\equiv \T{a}{c}\otimes \T{c}{b}$.  The
 $q$-generators $\chi$
are linear functionals on $G_q$. By taking
 the exterior derivative
of (\ref{qgenerators}), using $d^2=0$ and
the bi-invariance of $\tau=\ome{b}{b}$, we arrive
 at the $q$-Lie
algebra relations \cite{Jurco}, \cite{Aschieri}:
\eq
\cchi{d_1}{d_2} \cchi{c_1}{c_2} - \LL{e_1}{e_2}{f_1}{f_2}
{d_1}{d_2}{c_1}{c_2} ~\cchi{e_1}{e_2} \cchi{f_1}{f_2} =
\CC{d_1}{d_2}{c_1}{c_2}{a_1}{a_2} \cchi{a_1}{a_2}
\label{qLiealgebra}
\en
\noi where the structure constants are explicitly given by:
\eq
\CC{a_1}{a_2}{b_1}{b_2}{c_1}{c_2} =\lam [- \de^{b_1}_{b_2}
\de^{a_1}_{c_1}
\de^{c_2}_{a_2} +
\LL{b}{b}{c_1}{c_2}{a_1}{a_2}{b_1}{b_2}]. \label{CC}
\en
\noi and $\cchi{d_1}{d_2}
\cchi{c_1}{c_2} \equiv (\cchi{d_1}{d_2}  \otimes
\cchi{c_1}{c_2}) \D$.  Notice that
\eq
\LL{a_1}{a_2}{d_1}{d_2}{c_1}{c_2}{b_1}{b_2}=
\de^{b_1}_{a_1} \de^{a_2}_{b_2} \de^{c_1}_{d_1}
\de^{d_2}_{c_2} + O(q-\qm) \label{Lexpansion}
\en
because
the $R$ matrix itself has the form $R=I+(q-\qm)U$, with
$U$ finite in the $\qone$ limit,
see ref. \cite{FRT}. Then it is easy to see that (\ref{CC}) has
 a finite $\qone$ limit,
since  the $\lam$ terms cancel.
\sk
The Cartan-Maurer equations are found by
applying to $\ome{c_1}{c_2}$ the exterior
 differential as defined in
 (\ref{exteriordifferential}):
\eq
d\ome{c_1}{c_2}=\lam (\ome{b}{b} \we
 \ome{c_1}{c_2} + \ome{c_1}{c_2} \we
\ome{b}{b}) . \label{CartanMaurer}
\en
\noi Written as above, the Cartan-Maurer equations are
not of much use for
computations.  The right-hand side has an undefined
 ${0\over 0}$  classical limit.
 We need  a formula of the type
 $\ome{c_1}{c_2} \we \ome{b}{b}=
 -\ome{b}{b} \we\ome{c_1}{c_2}+ O(q-\qm)$
 that allows to eliminate in
(\ref{CartanMaurer}) the terms with the trace $\ome{b}{b}$
 (which has no classical counterpart) and obtain
an explicitly $\qone$ finite expression.
\sk
The desired ``$\om$ -permutator" can be found
 as follows. We first treat the
case of the $A_{n-1}$ series.
We apply relation (\ref{Hecke}) to the tensor
 product $\om \otimes \om$, i.e.:
\eq
(\L{ij}{kl}+q^2 \de^i_k \de^j_l)
(\L{kl}{mn}+q^{-2} \de^k_m \de^l_n)
(\L{mn}{rs}- \de^m_r \de^n_s)~\om^m
\otimes \om^n=0
\en
where we have used the adjoint
 indices ${~}^i \leftrightarrow {}_a^{~b}$,
 ${~}_i \leftrightarrow {}^a_{~b}$. Inserting the definition
of  the exterior product
 $\om^n \we \om^n=\om^m \otimes \om^n -\L{mn}{rs} \om^r \otimes \om^s$
yields
\eq
(\L{ij}{kl}+q^2 \de^i_k \de^j_l)(\L{kl}{mn}+
q^{-2} \de^k_m \de^l_n)~\om^m \we
\om^n = 0
\en
Multiplying by $\La^{-1}$ gives $(\La +(q^2+q^{-2})I+\La^{-1})~
\om\we\om$, or
equivalently
\eq
\om^i \we \om^j = - \Z{ij}{kl} \om^k \we \om^l \label{commom}
\en
\eq
\Z{ij}{kl} \equiv {1\over {q^2 + q^{-2}}} [\L{ij}{kl} + \Linv
{ij}{kl}]. \label{defZ}
\en
cf.  ref. \cite{Aschieri}. The $\om$-permutator
$\Z{ij}{kl}$ has the expected $\qone$ limit, that is $\de^i_l \de^j_k$.
\sk
There is another way to deduce the
 permutator $Z$, based on projector methods,
that we will use for the $B,C,D$ series. We first
 illustrate it in the easier $A$-case. Define
\eq
\PIJ{a_1}{a_2}{d_1}{d_2}{c_1}{c_2}{b_1}{b_2}
\equiv  d^{f_2} d^{-1}_{c_2}
\Rhat{b_1f_2}{c_2g_1} \PI{c_1g_1}{a_1e_1}
\Rhatinv{a_2e_1}{d_1g_2} \PJ{d_2g_2}{b_2f_2} \label{PIJ}
\en
with  {\small I,J=}$+,-$, the projectors $P_+,P_-$ being
 given in (\ref{projA}). The $(P_I,P_J)$ are themselves
 projectors, i.e.:
\eq
(P_I,P_J) (P_K,P_L) =\de_{IK} \de_{JL} (P_I,P_J) \label{projIJ}
\en
Moreover
\eq
(I,I)=I \label{IIeqI}
\en
so that
\eq
(I,I)=(P_++P_-,P_++P_-)=\Ppp + \Pmm +\Ppm +\Pmp=I \label{sumprojIJ}
\en
Eq.s (\ref{projIJ}) and (\ref{IIeqI}) are easy to prove by using
 (\ref{PIJ}) and the relation, valid for
all $A,B,C,D$ $q$-groups:
\eq
d^f d^{-1}_c \Rhat{bf}{cg} \Rhatinv{ce}{ba}=\de^f_a \de^e_g
\en
Projectors similar to (\ref{PIJ}) were already
introduced in ref. \cite{Watamura}.
{}From the definition (\ref{Lambda}) of $\La$  , using
 (\ref{RprojA})  and (\ref{PIJ})
we can write
\eq
\La=\Ppp+\Pmm-q^{-2} \Ppm-q^2 \Pmp  \label{Lproj}
\en
This decomposition shows that
$\La$ has eigenvalues
 $1,q^{\pm 2}$, and proves
 therefore eq. (\ref{Aspectral}).
{}From the definition of the exterior
product
$\om \we \om=\om \otimes \om-\La \om \otimes \om$
 we find the action of the projectors $(P_I,P_J)$ on $\om\we\om$:
\eq
\Ppp\om\we\om=\Pmm\om\we\om=0 \label{Pom}
\en
\eq
\Ppm\om\we\om=(1+q^{-2}) \Ppm\om\otimes\om,
{}~~\Pmp\om\we\om=(1+q^2)
\Pmp\om\otimes\om
\en
Using (\ref{sumprojIJ}) and (\ref{Pom}) we find :
\eq
\om \we \om=[\Ppp + \Pmm +\Ppm +
\Pmp] \om \we \om=[\Ppm +\Pmp] \om \we\om
\en
The $\om$-permutator is therefore
$Z=-\Ppm-\Pmp$. We can express it in
terms of the $\La$ matrix by observing that
\eq
\Ppm+\Pmp=-{1 \over q^2+q^{-2}}
 (\La+\La^{-1})+{2 \over q^2+q^{-2}}
(\Ppp+\Pmm) \label{ZprojA}
\en
as one deduces from (\ref{Lproj}). Note that $\La^{-1}$ is
 given in terms of projectors by the same expression as in
 (\ref{Lproj}), with $q \rightarrow \qm$. When acting on
 $\om\we\om$ the $\Ppp, \Pmm$ terms in (\ref{ZprojA})
can be dropped because of (\ref{Pom}), so
 that finally we arrive at eq. (\ref{defZ}).
\sk
Because of the expansion (\ref{Lexpansion}) and
a similar one for $\La^{-1}$ we
easily see that the $\om$-permutator (\ref{defZ}) can
be expanded as
\eq
\ZZ{c_1}{c_2}{d_1}{d_2}{a_1}{a_2}{b_1}{b_2}
\ome{a_1}{a_2} \we \ome{b_1}{b_2}=\ome{d_1}{d_2}
\we \ome{c_1}{c_2}+(q-\qm)
\WW{c_1}{c_2}{d_1}{d_2}{a_1}{a_2}{b_1}{b_2}
\ome{a_1}{a_2} \we \ome{b_1}{b_2} \label{Zlim}
\en
where $W$ is a finite matrix in the limit $\qone$.
\sk
Let us return to the Cartan-Maurer
 eqs. (\ref{CartanMaurer}).  Using (\ref{commom})
we can write:
\eq
d\ome{c_1}{c_2}=\lam (\ome{b}{b}
\we \ome{c_1}{c_2} -
\ZZ{c_1}{c_2}{b}{b}{a_1}{a_2}{b_1}{b_2}
\ome{a_1}{a_2} \we
\ome{b_1}{b_2}) \label{CartanMaurerZ}
\en
where $Z$ is given by
 $(\La+\La^{-1})/(q^2+q^{-2})$, cf.
(\ref{defZ}). Because
 of (\ref{Zlim}) we see that the
$\ome{b}{b}$ terms  disappear, and
(\ref{CartanMaurerZ}) has a finite
$\qone$ limit.
\sk
We now repeat the above construction for
the case of $q$-groups belonging to the $B,C,D$ series.
\sk
Using  (\ref{RprojBCD}) and (\ref{PIJ}) we find
the following projector decomposition for the $\La$ matrix :
\eq
\begin{array}{ll}
\La=&\Ppp+\Pmm+\Poo+
\epsi q^{\epsi-1-N}\Ppo+\epsi q^{-(\epsi-1-N)}\Pop\\
&-q^{-2} \Ppm-q^2 \Pmp -\epsi q^{N-\epsi-1}
\Pom-\epsi q^{-(N-\epsi-1)}\Pmo
\end{array}
\label{LprojBCD}
\en
from which we read off the eigenvalues of $\La$, and
prove eq. (\ref{BCDspectral}).
Proceeding as in the $A$ case, we find the action of
 the projectors on $\om\we\om$ :
\eq
\Ppp\om\we\om=\Pmm\om\we\om=\Poo \om\we\om=0 \label{PomBCD}
\en
\eqa
&\Ppm\om\we\om=(1+q^{-2})\Ppm\om\ot\om, \nonumber\\&\Pmp\om\we\om=
(1+q^2)\Pmp\om\ot\om
\ena
\eqa
&\Pmo\om\we\om=(1+\epsi q^{-(N-\epsi-1)})\Pmo\om\ot\om, \nonumber\\
&\Pom\om\we\om=(1+\epsi q^{N-\epsi-1})\Pom\om\ot\om  \label{Pom1BCD}
\ena
\eqa
&\Ppo\om\we\om=(1-\epsi q^{\epsi-1-N})\Ppo\om\ot\om, \nonumber\\
&\Pop\om\we\om=(1-\epsi q^{-(\epsi-1-N)})\Pop\om\ot\om \label{Pom2BCD}
\ena
Again the sum of the projectors $(P_I,P_J)$ yields the
 identity, so that we can write:
\eq
\om \we \om=[\Ppm +\Pmp+\Pmo+\Pom+\Ppo+\Pop] \om \we \om
\en
where we have taken (\ref{PomBCD}) into account. The
 $\om$-permutator $Z$  is therefore given by
\eq
Z=-[\Ppm +\Pmp+\Pmo+\Pom+\Ppo+\Pop]  \label{ZBCDproj}
\en
Can we express it in terms of odd
powers of  the $\La$ matrix, as in the case
of the $A$ groups ?
The answer is: only partially. In fact, by elementary
algebra we find
that
\eq
Z=-\al(\La+\La^{-1})-\be(\La^3+\La^{-3})
-(1-\al q_{-\epsi N}-\be q_{-3\epsi N})
[\Pso+\Pos] \label{ZBCD}
\en
with $\sigma \equiv sgn(\epsi)$ and
\eq
\al=-{{1+\be q_6} \over q_2}
\en
\eq
\be={{q_2-q_{\epsi N-2}}\over{q_6 q_{\epsi N-2}-q_2 q_{3(\epsi N-2)}}}
\en
\eq
q_n \equiv q^n+q^{-n}
\en
Note: $\La^r$ is given by
\eqa
&\La^r=\Ppp+\Pmm+\Poo
+\epsi^r [q^{r(\epsi-1-N)}\Ppo+ q^{-r(\epsi-1-N)}\Pop] \nonumber\\
&+(-1)^r[q^{-2r} \Ppm +q^{2r} \Pmp ] \nonumber\\
&+(-\epsi)^r [q^{r(N-\epsi-1)}\Pom+ q^{-r(N-\epsi-1)}\Pmo]
\ena
Let us check that $Z$ in (\ref{ZBCD}) has a
 correct classical limit. We have $\al \qonelim -
{9\over 16}$ and $\be \qonelim {1\over 16}$; taking into
account that
the $\Pso, \Pos$ terms disappear in the classical
 limit (cf. eq. (\ref{Pom1BCD}), (\ref{Pom2BCD}))
when applied to $\om\we\om$, we find the expected
 limit  $\Z{ij}{kl} \qonelim\de^i_l \de^j_k$.
\sk
The Cartan-Maurer equations are deduced
as before, and are given
by  (\ref{CartanMaurerZ}) where now $Z$ is the
 $\om$ permutator of eq. (\ref{ZBCD})
(Note: for explicit  calculations the expression
 (\ref{ZBCDproj}) or equivalently $Z=\Ppp+\Pmm+\Poo-I$
is more convenient).
Again the $\ome{b}{b}$ terms drop out
since $Z$ admits the expansion
$\ZZ{c_1}{c_2}{d_1}{d_2}{a_1}{a_2}{b_1}{b_2} \ome{a_1}{a_2}
 \we \ome{b_1}{b_2}=-\ome{d_1}{d_2}\we\ome{c_1}{c_2}+O(q-\qm)$.
\sk
In conclusion: we have found an
 explicit (and computable) expression
 for the Cartan-Maurer equations of
 the $\BCD$ $q$-groups. This opens the possibility
 of constructing gauge theories of these
$q$-groups,  following the procedure used
in \cite{Cas} for the $A_{n-1}$ $q$-groups.
\sk
Finally, let us comment on the differential calculus presented by the
authors of  ref.
\cite{Watamura}. Their definition of exterior
product in the $B,C,D$ case differs from
ours (and from the one adopted in \cite{Wor,Jurco,Zumino}), and
essentially amounts to
require that $\Pso \om \we \om=0$, $\Pos \om\we\om=0$, besides
(\ref{PomBCD}). This has one advantage: the term $\Pso+\Pos$ disappears
in the expression (\ref{ZBCD}). The disadvantage is that the defining
formula $\om\we\om=(I-\La)\om\ot\om$ does not hold any more for
the $B,C,D$ series, so that the general treatment of ref. \cite{Wor}
and the constructive procedure of ref.s \cite{Jurco} do not apply.

\vfill\eject

\vfill\eject
\end{document}